\documentclass[aps,prl,amsmath,amssymb,preprint]{revtex4-2}

\usepackage{braket}
\usepackage{amsmath}
\usepackage{graphicx}

\begin{document}

\title{Stimulated Raman lineshapes in the large light-matter interaction limit}

\author{%
	G. Batignani$^{1,2}$,
	G. Fumero$^{1}$,
	E. Mai$^{1}$,
	M. Martinati$^{1}$,
	T. Scopigno$^{1,2,3}$
}

\affiliation{$^{1}$Dipartimento di
	Fisica,~Universit\'a~di~Roma~``La Sapienza",~Roma,~I-00185,~Italy
	\email{tullio.scopigno@phys.uniroma1.it}
}
\affiliation{$^2$Istituto Italiano di Tecnologia, Center for Life Nano Science @Sapienza,~Viale Regina Elena, 291,~I-00161,~Italy
}
\affiliation{$^3$ Istituto Italiano di Tecnologia, Graphene Labs, Via Morego, 30, Genova,~I-16163, Italy
}

\maketitle

\section{abstract}
Stimulated Raman scattering (SRS) represents a powerful tool for accessing the vibrational properties of molecular compounds or solid state systems. 
From a spectroscopic perspective, SRS is able to capture Raman spectra free from incoherent background processes and typically ensures a signal enhancement of several orders of magnitude with respect to its spontaneous counterpart. Since its discovery in 1962, SRS has been applied to develop technological applications, such as Raman-based lasers, frequency shifters for pulsed sources and Raman amplifiers. 
For the full exploitation of their potential, however, it is crucial to have an accurate description of the SRS processes under the large gain regime.
Here, by taking as an example the stimulated Raman spectrum of a model solvent, namely liquid cyclohexane, we discuss how the spectral profiles and the lineshapes of Raman excitations critically depend on the pump excitation regime. In particular, we show that in the large light-matter interaction limit the Raman gain undergoes an exponential increase (decrease) in the red (blue) side of the spectrum, with the Raman linewidths that appear sharpened (broadened). 

\section{Introduction}

Stimulated Raman scattering was first reported in 1962, when Ng and Woodbury, gating a ruby laser for giant pulses production, detected an extra line shifted by an amount corresponding to the stretching mode of nitrobenzene, the material used for the Kerr shutter~\cite{woodbury1962ruby}. The observation was rationalized as the inelastic diffusion of (the Ruby emitted) light from a medium (nitrobenzene) coherently stimulated by the joint interaction of two propagating photons (the original Ruby 694.3nm photons and some Raman -Stokes- shifted ones), whose frequency difference matches one of the roto-vibrational frequencies of the material~\cite{Bloembergen1964}. Building on its coherent nature and chemical sensitivity, SRS has been widely exploited to investigate the structural properties and the photo-induced dynamics of molecules and solid state materials. In particular, the advent of ultrafast lasers has boosted novel spectro-microscopy approaches during the last thirty years~\cite{Yoshizawa1994,Ruhman1988,Kukura2005,Mathies_review,Kowalewski_2015,Prince2016,Polli2018,Batignani2019_CISRS,Fumero2020}.

In a typical spectroscopic configuration, SRS experiments require the presence of two external laser pulses temporally and spatially overlapped on the sample, namely a picosecond, narrowband ($\approx 10-20$ cm$^{-1}$) Raman pump (RP), which ensures high spectral resolution, and a femtosecond broadband probe pulse (PP)~\cite{Fang2009,Ernsting2010,Kuramochi2012,Quick2015,Batignani2015,Batignani2016,Dietze2016,Hall2017,Hontani2018,Otolski_2019,Batignani2019_adp,Ferrante2020}. 
Upon the system interaction with the RP, the PP impulsively stimulates vibrational coherences of the molecular modes, whose excitation energies correspond to the difference between the Raman and probe frequencies. 
The stimulated Raman signal is in fact induced by the third-order Raman susceptibility and is engraved on top of the PP, with the Raman modes detected as modifications of the PP spectral profile. Heterodyne detection hence efficiently suppresses any incoherent isotropic fluorescence background, and the net effect can be measured as the ratio of the probe spectral profile in presence and in absence of the Raman pump.
Under the electronically off-resonant condition, \textit{i.e.} when the incoming laser wavelengths are tuned far from any of the electronic absorption edges of the sample, higher energy RP photons are converted to lower energy photons in the red side spectral components of the PP. The opposite occurs in the blue side of the PP, where higher energy PP photons are converted to lower energy RP photons. For these reasons Raman gains and Raman losses are measured in the red and in the blue side of the SRS spectrum~\cite{Mukamel_2013}, respectively.
In parallel with microscopy and spectroscopy, technological applications based on SRS have been explored with the aim to realize Raman-based lasers ~\cite{benabid2002stimulated, Ferrara2020IntegratedRL, shen2020raman,adamu2021multi, wang2017demonstration}, frequency shifters~\cite{Vicario_2016, Guangyu_2021} and Raman amplifiers~\cite{Sirleto2020FiberAA, wang2014efficient}. 
Inelastic scattering of electromagnetic radiation provides indeed a convenient method to synthesize and amplify ultrashort laser pulses, which is in general a challenging task, with a unique versatility. The idea underneath Raman-based light emitting devices is to use the phonons response as the mediator for optical manipulation instead of electronic transitions, which are exploited in conventional laser technologies. For example, in Raman amplification, coherent phonons mediate the energy transfer from higher-frequency pump photons to lower-frequency signal photons~\cite{raymer1981stimulated}, similarly to the process discussed for the broadband probe in the SRS spectroscopic approach.
Taking advantage of Raman emission bands, which can be thousands of wavenumbers shifted with respect to the pump frequency, it is possible to switch toward different spectral regions. 
In this respect, since the first gaseous SRS laser demonstrated in 1963~\cite{minck1963laser}, multidisciplinary scientific efforts have been committed to the quest for ideal materials with optimal trade-off between large Raman gain and broad spectral range~\cite{Sirleto2012}.


For both spectroscopic investigations and the development of optical devices, it is important to understand  the mechanisms underlying an efficient generation of the SRS signal, which leads to large Raman gains. In particular, modifications in the lineshapes of the SRS spectra may reflect important details of the nonlinear process regardless of the specific application under consideration: in spectroscopy, lineshapes affect and possibly rule the spectral resolution, while in laser applications, spectral properties impacts directly on the generated bandwidth.

Since the SRS signal intensity is proportional to the incoming RP energy, the Raman gain ($RG$), defined as the ratio of the probe spectral profile in presence and in absence of the Raman pump, can be increased  varying the RP fluence. 
In the low RP energy fluence, $RG$ gain depends linearly on the pump intensity and can be calculated~\cite{cit::Mukamel,Dorfman2013} as $\left[ RG (\omega) -1 \right] \propto - \Im \left[P^{(3)}(\omega)/E_P(\omega)\right]$, where $\Im(z)$ denotes the imaginary part of $z$, $E_P(\omega)$ is the probe field and $P^{(3)}(\omega)$ is the third-order polarizability responsible for the SRS process. If the pump energy is increased beyond the linear threshold, this dependence is complicated by the nonlinear nature of the SRS process. 
Here we study the SRS response under nonlinear, high fluence pumping conditions. We measure the spectral modifications induced by a RP fluence beyond the linear regime of the $RG$ in cyclohexane, a liquid solvent used as a prototypical Raman scatterer. We interpret the experimental measurements by means of a microscopic treatment of the generated nonlinear polarization, showing that an overall sharpening of the Raman spectrum and a relative enhancement of the strongest Raman bands with respect to the weakest ones occur in the large light-matter interaction limit.

\section{Experimental Setup}

The Raman and probe pulses used for the SRS experiment are synthesized from the same source,  a Ti:sapphire laser, which generates transform limited 40 fs centered at 800 nm, with an energy of 3.2 mJ and 1 kHz of repetition rate.
In order to generate the PP, a small portion of the laser is focused on a sapphire 3 mm crystal producing a broadband (450-1000 nm) white light continuum (WLC) via supercontinuum generation~\cite{cit::Agrawal}. 
For the synthesis of the RP, a commercial two-stage optical parametric amplifier~\cite{Manzoni_2016} (Light Conversion TOPAS-C) is used to produce tunable IR-visible pulses, which are then frequency doubled by spectral compression (SC) via second harmonic and sum frequency generation ~\cite{Marangoni2007,Marangoni2009,Pontecorvo_2011} in a 25 mm-thick BBO crystal.
The picosecond Raman pulses generated by SC are centered at $\lambda_R \approx$ 580 nm and 
are characterized by a temporal profile unfavorable for SRS~\cite{Pontecorvo_2013,Hoffman2013}, which is then rectified by a double-pass (2f) spectral filter, further narrowing the pump spectrum. The RP fluence can be adjusted by a variable neutral density filter on the optical path.
The pulses, which are linearly polarized with parallel fields, are then focused on the sample in a non-collinear geometry ($\approx 5^\circ$) and the PP spectrum is monitored on a charge-coupled device (CCD) upon frequency dispersion by a spectrometer (Acton Spectra Pro 2500i).  In order to record the spectra of successive probe pulses in presence and in absence of the RP, a synchronized chopper blocks the RP at 500 Hz. Further details on the experimental setup are reported in~\cite{Pontecorvo_2011,Pontecorvo_2013,Batignani2020}.

\section{Results and Discussion}
\begin{figure}
	\includegraphics[width=0.9\textwidth]{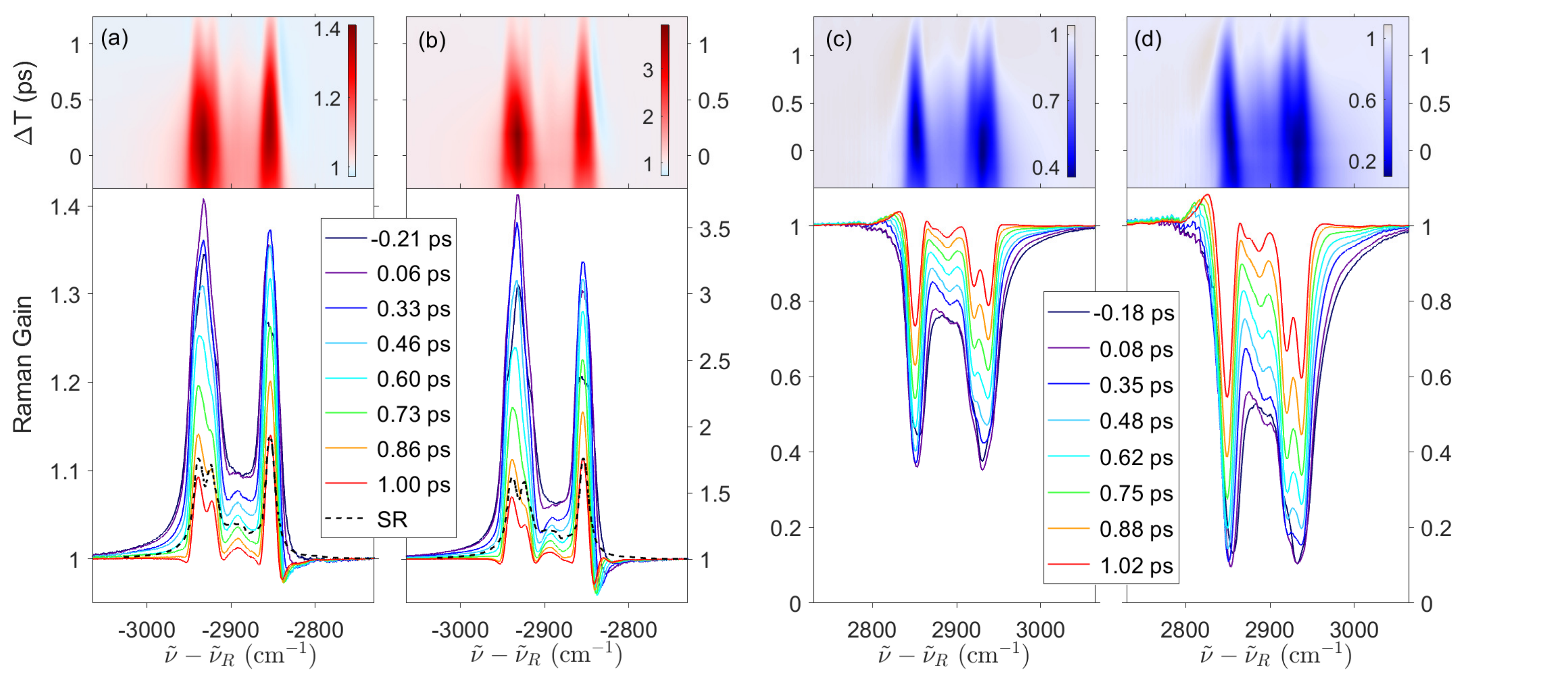}
	\caption{
    Red side SRS spectra of liquid cyclohexane acquired at two different fluences of the Raman pump (350 nJ in panel a and 1400 nJ  in panel b) measured varying the temporal delay between the RP and the PP. For positive time delays, the PP precedes the RP maximum providing an increased spectral resolution. 
    SRS spectra measured in the blue side are reported in panels c and d (with 500 nJ and a and 1400 nJ of RP excitation energy).
    Slices at selected time delays (indicated in ps in the legends) are reported in the bottom panels. The spontaneous Raman (SR) spectrum of cyclohexane, recorded with a continuous wave laser at 364 nm, is reported (in arbitrary units) in panels (a)-(b) for comparison~\cite{bell1998analysis}.}
	\label{Fig:data}
\end{figure}

\begin{figure}
	\centering
		\includegraphics[scale=.5]{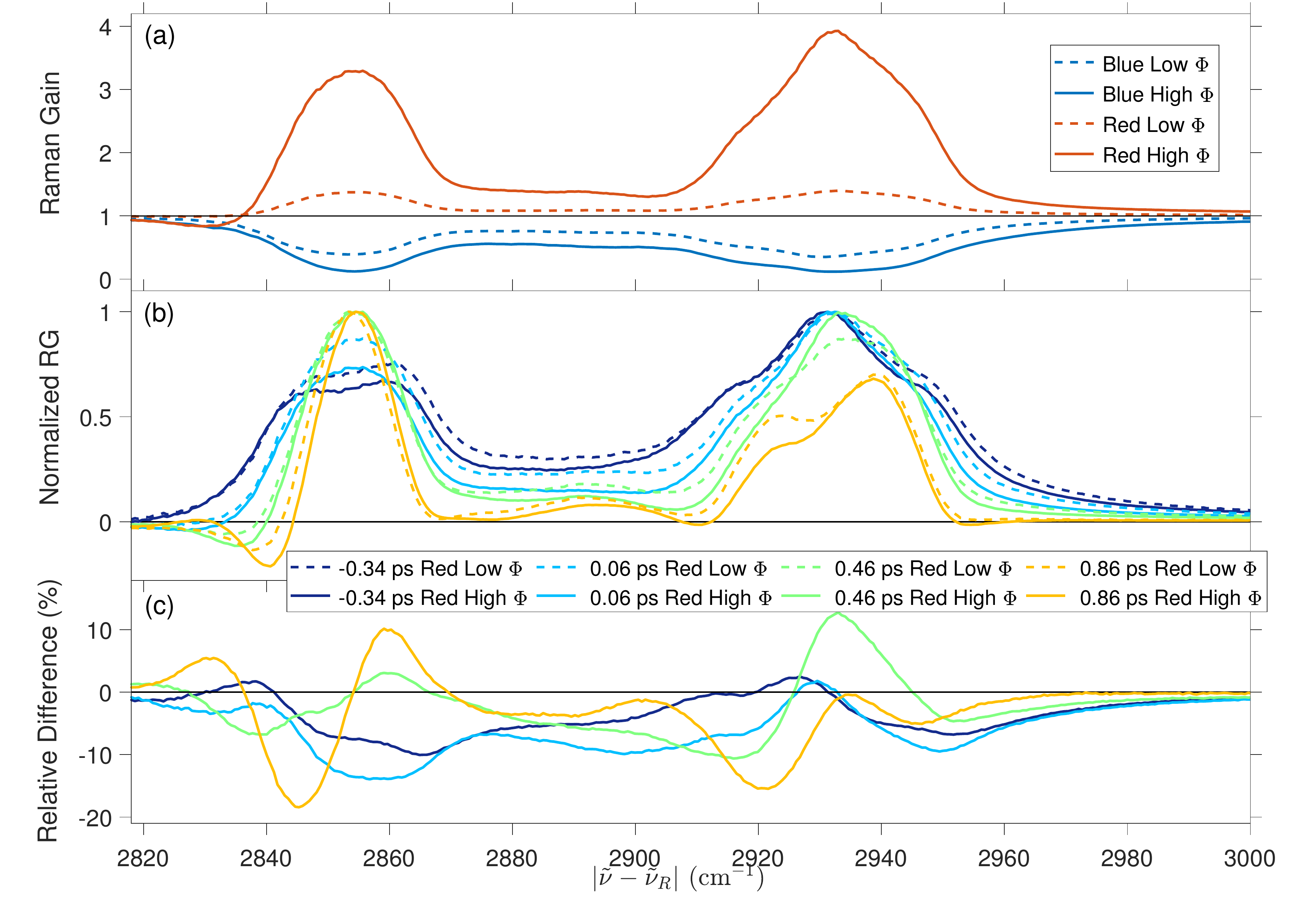}
	\caption{
    SRS spectra acquired at low and high fluences ($\Phi$) in the red and in the blue side of the spectrum are compared in panel (a) as a function of the absolute Raman shift for overlapped Raman and probe pulses.
    The Raman gain normalized to the corresponding maximum is reported in panel (b) for selected time delays reported in the legend in ps at high and low fluences (continuous and dashed lines). Notably, the spectral lineshapes and the relative peak intensities show pronounced differences varying the fluence regime, with the strongest Raman peak that undergoes an enhancement and a sharpening with respect to the weakest ones. This effect is emphasized in panel (c), where we report the difference between the normalized Raman spectra of panel (b) acquired at high vs low fluence.}
	\label{Fig:Compare_data}
\end{figure}

SRS measurements have been performed on a liquid solvent, namely cyclohexane ($\mathrm{C_6H_{12}}$),  and are reported in Figure~\ref{Fig:data} as a function of the temporal delay $\Delta T$ between the Raman and the probe pulses for different excitation RP fluences. The spontaneous Raman spectrum of cyclohexane ~\cite{bell1998analysis} is also reported for comparison.
In order to investigate the SRS lineshape in the large light-matter interaction limit, 
we focused on the high frequency spectral region, where the large cross-section of the $\mathrm{C-H}$ bond vibrations generates strong Raman signals.  In addition, the overlapping Raman bands around 2900 cm$^{-1}$ offer the chance to investigate the dependence of the spectral resolution as a function of the Raman pump fluence.
In order to avoid cross phase modulation artifacts~\cite{cit::Agrawal,Lim2018,Batignani2019}, we limited the RP energy below 2 $\mu J$, using a long, 10 mm-thick, optical glass cuvette to increase the SRS response.
The SRS spectra collected for PP spectral components  red shifted with respect to the RP wavelength ($\tilde{\nu}<\tilde{\nu}_R$) are reported in panels (a-b), while the blue side spectra ($\tilde{\nu}>\tilde{\nu}_R$) are shown in panels (c-d). As discussed in Ref.~\cite{Mukamel_2013}, at odds with the spontaneous Raman case, the blue side spectra are originated from initial molecular populations in the ground vibrational state and hence the amplitude of the high frequency mode does not vanish at room temperature. A comparison of the red and the blue side spectra is reported in Figure~\ref{Fig:Compare_data}-a.
As expected, the Raman gain profiles reach the maximum intensity for slightly positive time delays ($\Delta T > 0$), \textit{i.e.} for a probe pulse preceding the RP maximum, in agreement with previous observations~\cite{Yoon2005,Ferrante2018}.

\begin{figure}
	\centering
	\includegraphics[scale=.38]{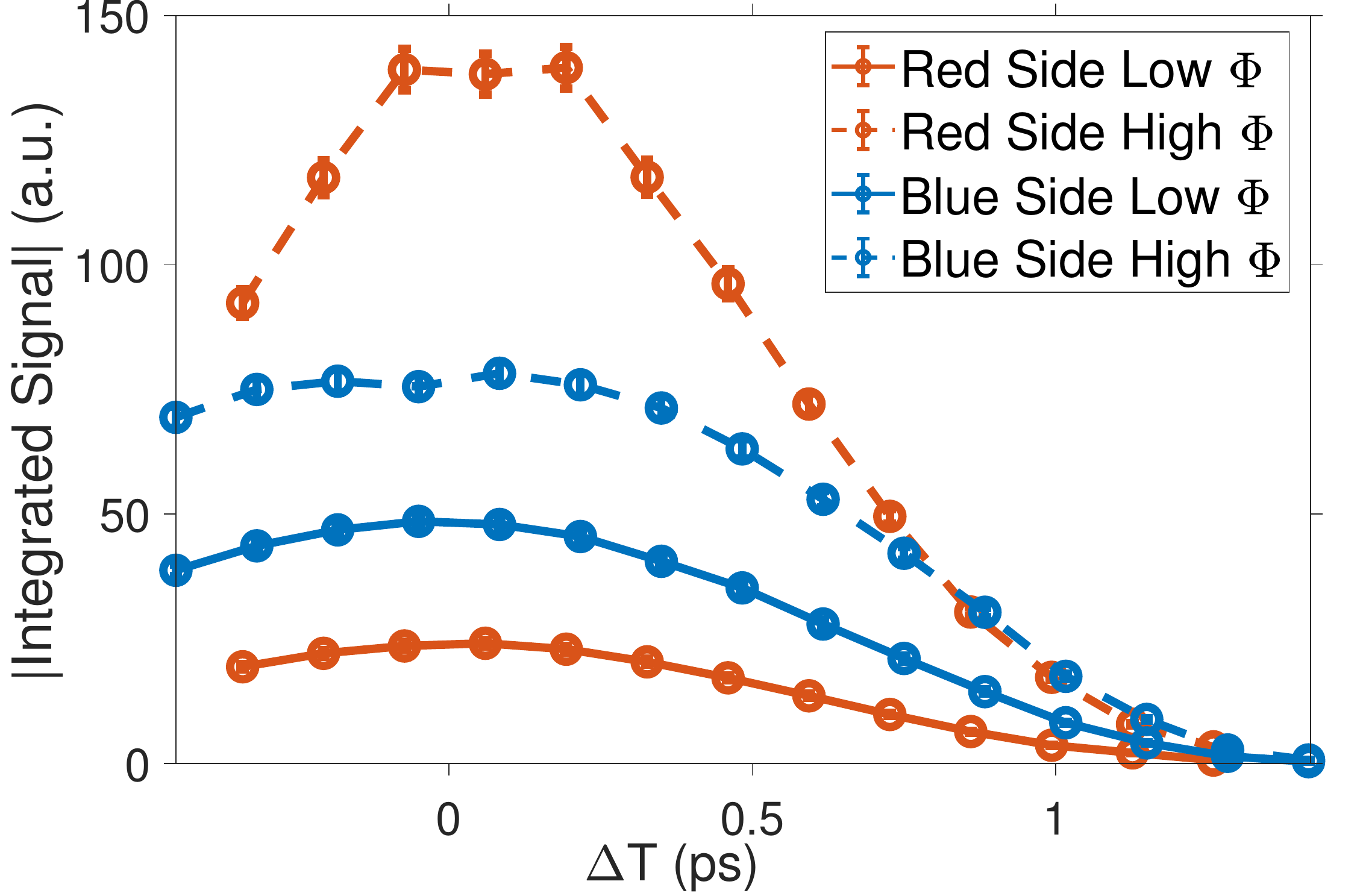}
	\caption{
		Integrated area of the measured SRS signal as a function of the RP-PP delay in the red and in the blue side of the spectrum for the low and high fluence regimes.}
	\label{Fig:Area}
\end{figure}

Notably, for the RP-PP temporal delays that provide the maximum Raman gain, increasing the RP energy by a factor of 4 (from 350 nJ to 1400 nJ) results in red-side Raman bands with an amplitude $\approx$ 7 times larger, while an increase of the RP energy by a factor of $\approx$ 3 (from 500 nJ to 1400 nJ) generates blue-side Raman loss with relative amplitudes that increase by a $\approx$ 1.4 factor.
In addition, a careful inspection of Figure \ref{Fig:data} reveals  different spectral lineshapes and relative peak intensities for different pump fluences.
This is emphasized in Figure \ref{Fig:Compare_data}-b, where the red side SRS spectra normalized to their corresponding maximum are shown for selected time delays at low and high RP fluences (dashed and continuous lines, respectively). Specifically, the strongest Raman peak at 2938 cm$^{-1}$ undergoes an enhancement and a sharpening with respect to the weaker 2924 cm$^{-1}$ neighboring mode, which can be barely resolved in the high fluence regime. The difference between normalized SRS spectra, shown in Figure \ref{Fig:Compare_data}-c, further stresses the significant difference (up to 20\%) between the low and the high fluence spectral profiles.  In Figure \ref{Fig:Area}, the integrated area of the SRS profiles as a function of the RP-PP delay are shown for the red and the blue side.

In order to rationalize the peculiar behavior of such experimental observations, we modeled the SRS response 
by a perturbative expansion of the light-matter interaction in terms of the electromagnetic fields.
Starting from the expression of the Maxwell's equations inside a dielectric medium, it is readily obtained in Gaussian units~\cite{cit::Mukamel}
\begin{equation}\label{Eq:Max}
    \nabla \times \nabla \times E(\mathrm{r},t) + \frac{n^2}{c^2}\frac{\partial^2 E(\mathrm{r},t)}{\partial t^2}=-\frac{4\pi}{c^2}\frac{\partial^2 P^{(3)}(\mathrm{r},t)}{\partial t^2}
\end{equation}
where $P^{(3)}$ is the third-order nonlinear polarization responsible for the SRS process, while the corresponding term for the linear polarization has been included in the refractive index $n$.
The RP and PP fields can be conveniently expressed in terms of their temporal envelopes $\mathcal{A}_R(z,t)$ and $\mathcal{A}_P(z,t)$ as  $E_{R/P}(z,t)=\mathcal{A}_{R/P}(z,t) \, e^{i(k_{R/P}\,z - \omega_{R/P}\,t)}+c.c.$, where $\hat{z}$ is the propagation direction and $\omega_{R/P}$ is the carrier frequency of the Raman/Probe pulse, respectively. Analogously, the nonlinear polarization in the $R/P$ field frame can be written as $P^{(3)}_{R/P}(z,t) = \mathcal{P}_{R/P}^{(3)}(z,t) \, e^{i(k_{R/P}\, z - \omega_{R/P}\, t)} + c.c.$ Under the slowly varying envelope approximation (SVEA), namely if $\frac{\partial^2}{\partial t^2}\mathcal{A}_{R/P}(z,t) \ll \omega_{R/P} \, \frac{\partial}{\partial t}\mathcal{A}_{R/P}(z,t) \ll \omega^2_{R/P} \, \mathcal{A}_{R/P}(z,t)$ and $\frac{\partial^2}{\partial t^2}\mathcal{P}^{(3)}_{R/P}(z,t) \ll \omega_{R/P} \, \frac{\partial}{\partial t}\mathcal{P}^{(3)}_{R/P}(z,t)  \ll \omega^{2}_{R/P} \, \mathcal{P}^{(3)}_{R/P}(z,t)$, the second derivative of $\mathcal{A}_{R/P}(z,t)$ with respect to $z$ as well as the first and second derivatives with respect to time of the two envelopes in Eq. \ref{Eq:Max} can be neglected. As a result, it holds
\begin{equation}\label{Eq:parabolic}
    ik_{R/P}\frac{\partial \mathcal{A}_{R/P}(t)}{\partial z}=- \frac{2\pi}{c^2}\omega_{R/P}^2 \, \mathcal{P}^{(3)}_{R/P}(t) \, e^{i \mathrm{\Delta k} \cdot \mathrm{z}}
\end{equation}


\noindent
where $ \Delta k$ is the wavevector mismatch~\cite{Boyd_book}. 
The last equation can be expressed in the frequency domain as
\begin{equation}\label{Eq:parabolic_fd}
    ik_{R/P}\frac{\partial \mathcal{A}_{R/P}(\omega)}{\partial z}=- \frac{2\pi}{c^2}\omega_{R/P}^2 \, \mathcal{P}^{(3)}_{R/P}(\omega) \,e^{i \mathrm{\Delta k} \cdot \mathrm{z}}
\end{equation}

\noindent
and it holds $\mathcal{A}_{R/P}(\omega) = E_{R/P}(\omega +\omega_{R/P})$ as well as $\mathcal{P}^{(3)}_{R/P}(\omega) = P^{(3)}_{R/P}(\omega + \omega_{R/P})$.

\begin{figure}
	\centering
		\includegraphics[scale=.5]{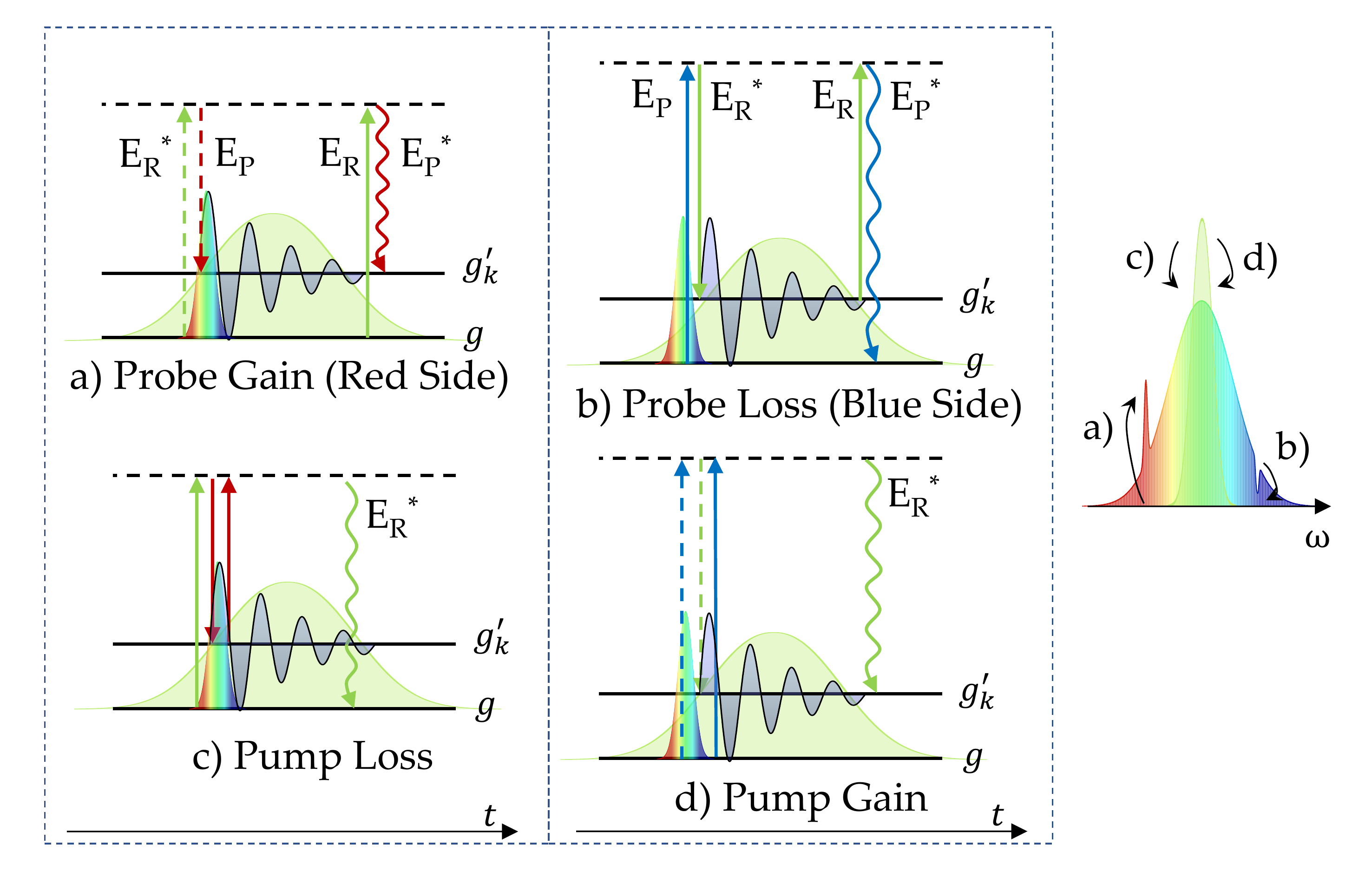}
	\caption{
	Energy level diagrams accounting for SRS gains in the red and losses in the blue side of the probe spectrum (a and b, respectively). Two interactions with the Raman and probe fields stimulate a vibrational coherence, which is then probed by a subsequent interaction with the Raman field and a free-induction decay. Diagrams accounting for loss and gains on the Raman pump are depicted in the bottom panels (c and d, respectively). 
	Labels $g$, $g'_k$ denote the ground state and a vibrationally excited level.
	Dashed and continuous arrows are used to indicate interactions with the bra and ket sides of the density matrix, respectively.}
	\label{Fig:diagrams}
\end{figure}

\vspace{0.15cm}
It is worth to stress that, in principle, the third-order nonlinear polarization $P^{(3)}$ that generates the Raman signal results from different processes, corresponding to the different permutations of the fields-matter interactions~\cite{Dorfman2013,Lee2004,Batignani2015pccp,Batignani_2021}. Under the non-resonant regime considered in the present work, the concurring processes, responsible for the SRS gains and losses in the Raman and probe fields, are depicted in Figure~\ref{Fig:diagrams}. Briefly, diagrams a-b describe the stimulation of a vibrational coherence by two consecutive interactions between the sample and the RP and PP fields, followed by a second interaction with the RP and a free induction decay.
Similarly, diagrams c-d describe a SRS Raman loss/gain in the RP field, generated by a double interaction with the red/blue shifted spectral components of the PP.
Focusing on the SRS signal generated on the probe field, the third-order polarization originated by the a-b diagrams can be expressed in the frequency domain as~\cite{cit::Mukamel}
\begin{multline}
    \label{Eq:Pta}
    P^{(3)}_a(\omega)=\left(\frac{i}{\hbar}\right)^3 \sum_k
    \int_{-\infty}^{\infty} dt \,e^{i\omega t}
    \int_0^{+\infty}d\tau_3 E_R(t-\tau_3)\\
    \int_0^{+\infty}d\tau_2 E_P(t-\tau_2-\tau_3)
    \int_0^{+\infty}d\tau_1
    E^*_R(t-\tau_1-\tau_2-\tau_3)
    \\
    |\mu_{ge}|^2
    |\mu_{eg'_k}|^2
    e^{-i\tilde{\omega}_{ge}\tau_1}
    e^{-i\tilde{\omega}_{gg'_k}\tau_2} 2\pi\,g(\tau_2)
    e^{-i\tilde{\omega}_{eg'_k}\tau_3}
\end{multline}
and 
\begin{multline}
    \label{Eq:Ptb}
    P^{(3)}_b(\omega)=\left(\frac{i}{\hbar}\right)^3 \sum_k
    \int_{-\infty}^{\infty} dt \,e^{i\omega t}
    \int_0^{+\infty}d\tau_3 E_R(t-\tau_3)\\
    \int_0^{+\infty}d\tau_2 E_R^*(t-\tau_2-\tau_3) \int_0^{+\infty}d\tau_1 E_P(t-\tau_1-\tau_2-\tau_3)
    \\ 
    |\mu_{ge}|^2
    |\mu_{eg'_k}|^2
    e^{-i\tilde{\omega}_{eg}\tau_1}
    e^{-i\tilde{\omega}_{g'_kg}\tau_2} 2\pi\,g(\tau_2)
    e^{-i\tilde{\omega}_{eg}\tau_3}
\end{multline}
where the summation over $k$ runs over the different investigated Raman active modes; $\tau_i$ represents a time interval between two consecutive radiation-matter interactions, 
$\tilde{\omega}_{ij}=\omega_{i}-\omega_{j}-i\gamma_{ij}$, with  $\gamma_{ij}=\tau_{ij}^{-1}$ that indicates the dephasing time of the $\ket{i}\bra{j}$ coherence, $\mu_{ij}$ is the dipole matrix element between the $i$-$j$ states, and $2\pi\,g(\tau)$ is an inhomogeneous dephasing function additional to the natural exponential damping (which can be neglected in the homogeneous regime). The $g$, $g'_k$, $e$ labels indicate the molecular ground state, the vibrationally excited and the off-resonant electronically excited levels, respectively. Importantly, the SRS signal is generated collinearly with the probe pulse direction ($k_P=k_P+k_R-k_R^*$) and the process is self phase-matched ($\Delta k=0$).

By recasting the RP/PP envelopes and the inhomogeneous dephasing function in terms of their Fourier transform 
$$E_{R/P}(t)=\int_{-\infty}^{\infty} \frac{d\omega}{2 \pi} \, E_{R/P}(\omega) \, e^{-i\omega t} 
$$ 
$$g(\tau) = \int_{-\infty}^{\infty} \frac{d\omega_D}{2 \pi} \, g(\omega_D) \, e^{-i\omega_D t}$$
it is possible to simplify Eqs. \ref{Eq:Pta}, \ref{Eq:Ptb} to

\begin{equation}
	\begin{gathered}
    \label{Eq:Pta2}
    P^{(3)}_a(\omega)=\left(\frac{i}{\hbar}\right)^3 \sum_k     |\mu_{ge}|^2
    |\mu_{eg'_k}|^2 \, \frac{1}{(2 \pi)^3} \int_{-\infty}^{\infty} d\omega_D \, g(\omega_D) \\
     \int_{-\infty}^{\infty} d\omega_1 
     E^*_R(\omega_1)
    \int_{-\infty}^{\infty} d\omega_2 
    E_P(\omega_2) 
    \int_{-\infty}^{\infty} d\omega_3  
    E_R(\omega_3) \\
    \int_{-\infty}^{\infty} dt \,e^{i\omega t}
    \int_0^{+\infty}d\tau_3
    \int_0^{+\infty}d\tau_2 
    \int_0^{+\infty}d\tau_1\\
    e^{i\omega_1(t-\tau_1-\tau_2-\tau_3)}
    e^{-i\omega_2(t-\tau_2-\tau_3)}
    e^{-i\omega_3(t-\tau_3)}
    \\
    e^{-i\tilde{\omega}_{ge}\tau_1}
    e^{-i(\tilde{\omega}_{gg'_k}+\omega_D)\tau_2}
    e^{-i\tilde{\omega}_{eg'_k}\tau_3}
	\end{gathered}
\end{equation}
and
\begin{equation}
	\begin{gathered}
    \label{Eq:Ptb_2}
    P^{(3)}_b(\omega)=\left(\frac{i}{\hbar}\right)^3 \sum_k 
    |\mu_{ge}|^2
    |\mu_{eg'_k}|^2  \, \frac{1}{(2 \pi)^3} \int_{-\infty}^{\infty} d\omega_D\, g(\omega_D) \\
     \int_{-\infty}^{\infty} d\omega_1  E_P(\omega_1)
      \int_{-\infty}^{\infty} d\omega_2 E_R^*(\omega_2)
       \int_{-\infty}^{\infty} d\omega_3 E_R(\omega_3)
       \\
    \int_{-\infty}^{\infty} dt \,e^{i\omega t}
    \int_0^{+\infty}d\tau_3 \int_0^{+\infty}d\tau_2\int_0^{+\infty}d\tau_1 \\
    e^{-i\omega_3(t-\tau_3)}
     e^{i\omega_2(t-\tau_2-\tau_3)}  e^{-i\omega_1(t-\tau_1-\tau_2-\tau_3)}
    \\ 
    e^{-i\tilde{\omega}_{eg}\tau_1}
    e^{-i(\tilde{\omega}_{g'_kg}+\omega_D)\tau_2}
    e^{-i\tilde{\omega}_{eg}\tau_3}
	\end{gathered}
\end{equation}
where all the temporal integrals can be analytically simplified as follows: 
\begin{equation}
\begin{gathered}
    \label{Eq:Pta3}
    P^{(3)}_a(\omega)=\left(\frac{1}{\hbar}\right)^3 \sum_k     |\mu_{ge}|^2
    |\mu_{eg'_k}|^2 \, \frac{1}{(2\pi)^2}
    \int_{-\infty}^{\infty} d\omega_D \, g(\omega_D)\\ 
     \int_{-\infty}^{\infty} d\omega_1 
     E^*_R(\omega_1) 
    \int_{-\infty}^{\infty} d\omega_2 
    E_P(\omega_2) 
    \int_{-\infty}^{\infty} d\omega_3 
    E_R(\omega_3)\\
    \frac{\delta(\omega+\omega_1-\omega_2-\omega_3)}{(\tilde{\omega}_{eg'_k}+\omega_1-\omega_2-\omega_3)(\tilde{\omega}_{gg'_k}+\omega_1-\omega_2)(\tilde{\omega}_{ge}+\omega_1)}
\end{gathered}
\end{equation}
and
\begin{equation}
\begin{gathered}
    \label{Eq:Ptb_3}
    P^{(3)}_b(\omega)=\left(\frac{1}{\hbar}\right)^3 \sum_k 
    |\mu_{ge}|^2
    |\mu_{eg'_k}|^2 \, \frac{1}{(2 \pi)^2}
    \int_{-\infty}^{\infty} d\omega_D \, g(\omega_D)\\ 
     \int_{-\infty}^{\infty} d\omega_1  E_P(\omega_1)
      \int_{-\infty}^{\infty} d\omega_2 E_R^*(\omega_2)
       \int_{-\infty}^{\infty} d\omega_3 E_R(\omega_3)\\
       \frac{\delta(\omega-\omega_3+\omega_2-\omega_1)}{(\tilde{\omega}_{eg}-\omega_3+\omega_2-\omega_1)(\tilde{\omega}_{g'_kg}+\omega_2-\omega_1)(\tilde{\omega}_{eg}-\omega_1)}
\end{gathered}
\end{equation}
These last equations can be recast by taking advantage of the Dirac delta, simplifying the integral over the probe field, obtaining
\begin{equation}
\begin{gathered}
    \label{Eq:Pta4}
    P^{(3)}_a(\omega)=\sum_k \frac{|\mu_{ge}|^2 |\mu_{eg'_k}|^2 }{(2\pi)^2 \hbar^3}      
     \int_{-\infty}^{\infty}  \int_{-\infty}^{\infty} \int_{-\infty}^{\infty}d\omega_D\,d\omega_1 
     d\omega_3 \\ 
    \frac{g(\omega_D) E_R(\omega_3) E^*_R(\omega_1)E_P(\omega+\omega_1-\omega_3) }{(\tilde{\omega}_{eg'_k}-\omega)(\tilde{\omega}_{gg'_k}-\omega+\omega_3)(\tilde{\omega}_{ge}+\omega_1)}
\end{gathered}
\end{equation}
and
\begin{equation}
\begin{gathered}
    \label{Eq:Ptb_4}
    P^{(3)}_b(\omega)=\sum_k\frac{ |\mu_{ge}|^2
    |\mu_{eg'_k}|^2}{(2\pi)^2 \hbar^3}  
    \int_{-\infty}^{\infty} \int_{-\infty}^{\infty} \int_{-\infty}^{\infty}
    d\omega_D \, d\omega_2 d\omega_3\\
     \frac{g(\omega_D) E_P(\omega-\omega_3+\omega_2)
     E_R^*(\omega_2)
     E_R(\omega_3)}{(\tilde{\omega}_{eg}-\omega)
     (\tilde{\omega}_{g'_kg}-\omega+\omega_3)(\tilde{\omega}_{eg}-\omega-\omega_2+\omega_3)}
\end{gathered}
\end{equation}
that are the convolution of the third-order polarizations $P^{(3)}_{a/b}(\omega)$ valid for the homogeneous lineshape regime with the additional dephasing function $g(\omega_D)$.

Considering a monochromatic Raman pulse, $E_R(t)=\mathcal{A}_R^{0}e^{-i\omega_R t}$, $E_R(\omega)= \mathcal{A}_R^{0} \, 2 \pi \delta(\omega-\omega_R)$ and neglecting hereafter the inhomogeneous dephasing ($g(\omega_D)=\delta(\omega_D)$), the integral in Eqs. \ref{Eq:Pta4} and \ref{Eq:Ptb_4} can be simplified, leading to the following analytical relations
\begin{equation}
\begin{gathered}
    \label{Eq:Pta5}
    P^{(3)}_a(\omega)= \sum_k     
    \frac{ \hbar^{-3} \, |\mu_{ge}|^2 |\mu_{eg'_k}|^2 \, |\mathcal{A}_R^{0}|^2 \, E_P(\omega) }{(\tilde{\omega}_{eg'_k}-\omega)
    (\tilde{\omega}_{gg'_k}-\omega+\omega_R)
    (\tilde{\omega}_{ge}+\omega_R)}=\\
    \chi^{(3)}_a(\omega_R,\omega) |\mathcal{A}_R^{0}|^2E_P(\omega)
\end{gathered}
\end{equation}
and
\begin{equation}
	\begin{gathered}
    \label{Eq:Ptb_5}
    P^{(3)}_b(\omega)=\sum_k 
     \frac{ \hbar^{-3} \, |\mu_{ge}|^2
    |\mu_{eg'_k}|^2 \, |\mathcal{A}_R^{0}|^2 \, E_P(\omega)
     }{(\tilde{\omega}_{eg}-\omega)
     (\tilde{\omega}_{g'_kg}-\omega+\omega_R)(\tilde{\omega}_{eg}-\omega)}=\\
      \chi^{(3)}_b(\omega_R,\omega) |\mathcal{A}_R^{0}|^2E_P(\omega)
\end{gathered}
\end{equation}

Combining Eqs. \ref{Eq:parabolic_fd} and \ref{Eq:Pta5}, it is possible to calculate the spectrally dispersed SRS response in the red/blue side of the spectrum, accordingly to
\begin{equation}\label{Eq:parabolic_fd_2}
    ik_{P}\frac{\partial \mathcal{A}_{P}(z, \omega)}{\partial z}=- \frac{2\pi}{c^2}\omega_{P}^2 
    \chi^{(3)}_{a/b}(\omega_R,\omega) |\mathcal{A}_R^{0}|^2 \mathcal{A}_{P}(z, \omega)
\end{equation}
that can be integrated analytically:
\begin{equation}\label{Eq:parabolic_fintegrated}
    \begin{cases}
    \mathcal{A}_{P}(z,\omega)=\mathcal{A}_{P}(0,\omega) \, e^{+i \frac{2\pi \omega_{P} \chi^{(3)}_{a/b}(\omega_R,\omega) I_Rz}{c \,n_{P} }}
    \\
    
    I_{P}(z,\omega+\omega_{P})=I_{P}(0,\omega+\omega_{P}) \, e^{-\frac{2\pi \omega_{P} \Im \left[\chi^{(3)}_{a/b}(\omega_R,\omega) \right] I_R z}{c n_{P} }}
    \end{cases}
\end{equation}
Hence, the measured Raman gain $RG(\omega)=\frac{I_P^{R_{on}}(\omega)}{I_P^{R_{of\!f}}(\omega)}=\frac{I_{P}(z, \, \omega)}{I_{P}(0, \, \omega)}$ is equal to
\begin{equation}\label{eq: RG_z}
RG_{\hbox{red/blue}}(\omega)=e^{-\frac{2\pi \omega_{P} \Im \left[\chi^{(3)}_{a/b}(\omega_R, \,\omega - \omega_P) \right] I_R z}{c n_{P} }}
\end{equation}
which, for small gains, can be expanded to the first order in $I_R z$ as
$
RG_{\hbox{red/blue}}(\omega)\approx 1-\frac{2\pi \omega_{P} \Im \left[\chi^{(3)}_{a/b}(\omega_R,\omega - \omega_P) \right] I_R z}{c n_{P} }
$
, so that it depends linearly on the imaginary part of $\chi^{(3)}_{a/b}$. We note that in the low excitation regime, the $\mathcal{A}_{P}(z,\omega)$ in the right hand side of Eq. \ref{Eq:parabolic_fd_2} can be considered constant, retrieving the  Raman Gain definition reported in the introduction for low RP excitation regimes.

It is worth to stress that, under the low RP excitation regime, the linear dependence of the RG on the probed frequency $\omega_P$ in Eq. \ref{eq: RG_z} generates blue side spectra with an amplitude stronger with respect to the red side.  
Notably, while in the red side $-\Im \left( \chi^{(3)}_{a}\right)$ corresponds to the sum of positive Raman bands and hence generates signals that exponentially grow as a function of the sample length and RP intensity, in the blue side $-\Im \left( \chi^{(3)}_{b}\right)$ is a negative quantity, leading to exponentially decaying Raman gains.
\begin{figure}
	\centering
		\includegraphics[scale=.275]{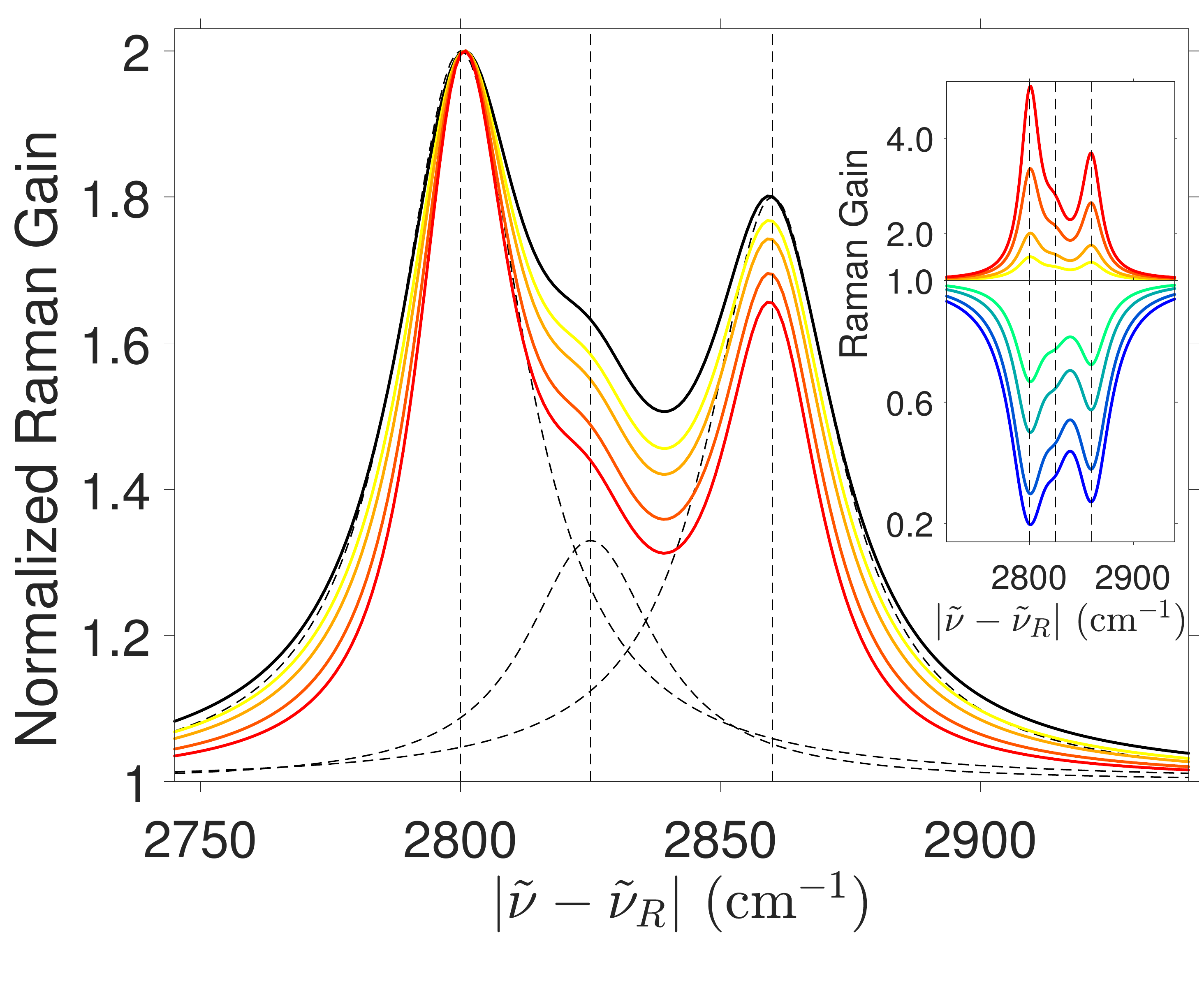}
	\caption{Normalized red-side SRS lineshapes obtained from Eq. \ref{eq: RG_z} varying the fluence of a monochromatic RP (the corresponding relative amplitudes are 1.0, 1.7, 3.0, 4.0), calculated considering three Raman modes (at 2800, 2825 and 2860 cm$^{-1}$) with different amplitudes and 30 cm$^{-1}$ full width half maxima. The black continuous line indicates the spontaneous Raman spectrum, while the dashed lines represent the different Raman  components. The inset shows the corresponding Raman gain in the the red and in the blue side of the spectrum.
    }
	\label{Fig:sketch_lineshapes}
\end{figure}
As illustrated in Figure \ref{Fig:sketch_lineshapes}, such exponential gain alters the relative intensity and the lineshapes of the different modes. Specifically, the strongest Raman modes are enhanced with respect to the weakest ones, and the overall Raman profile is sharpened. 


\begin{figure}
	\centering
	\includegraphics[scale=.405]{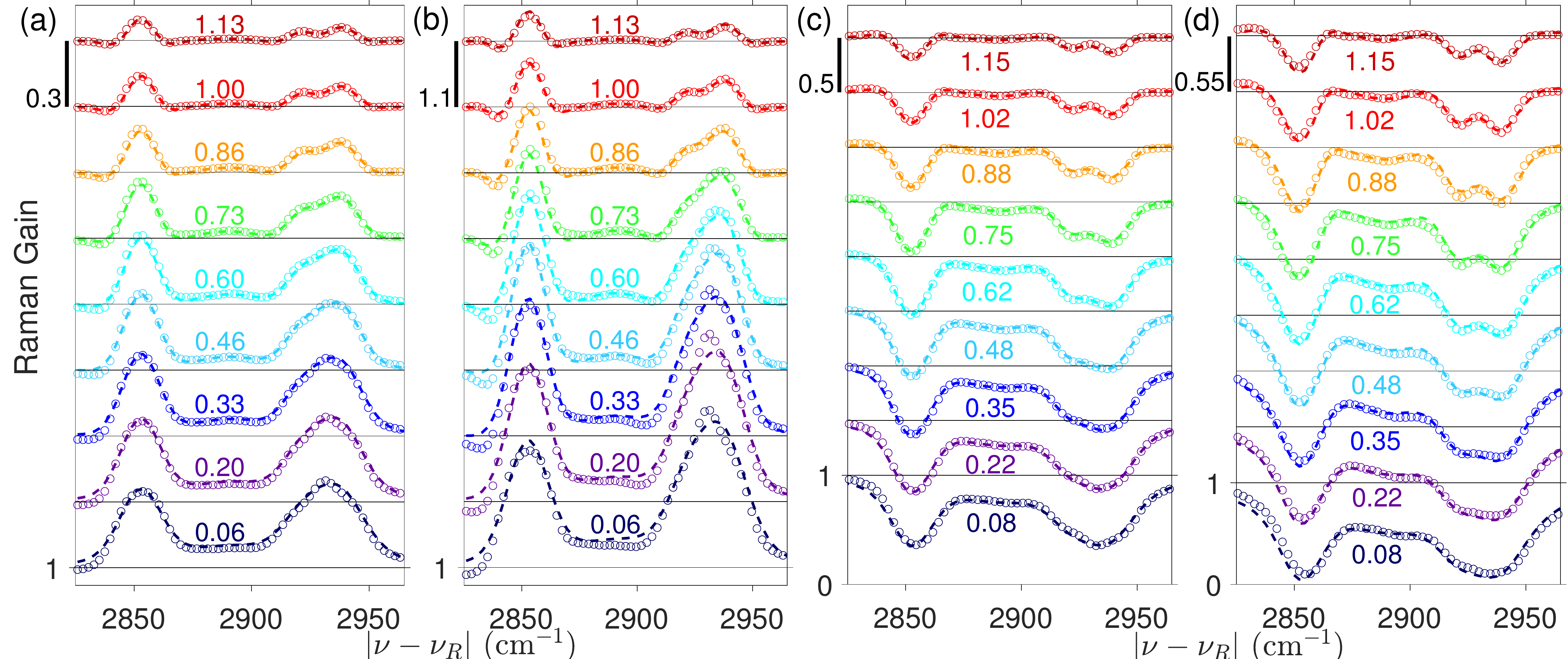}
	\caption{
   Experimental cyclohexane SRS spectra (circles) are compared with the modeled ones (dashed lines) in the red side of the spectrum under the low and the high excitation regimes (panels a and b, corresponding to a 350 nJ and 1400 nJ Raman pump energy, respectively) and in the blue side (panels c and d, measured with a 500 nJ and 1400 nJ RP energy). Traces acquired at different temporal delays between the Raman and the probe fields (indicated in the legends) have been vertically offset by a constant factor.}
	\label{Fig:Fit_red}
\end{figure}

For non-monochromatic Raman pump pulses, Eq. \ref{eq: RG_z} is no longer valid to calculate the SRS spectrum, which hence has to be computed by numerically propagating the probe field by Eq. \ref{Eq:parabolic_fd} coupled to Eqs. \ref{Eq:Pta4}-\ref{Eq:Ptb_4}.
Since the PP is itself acting as a source of the $P^{(3)}_{a/b}(\omega)$, this can be conveniently done by a symmetrized split-step method in which numerical integrations are performed at small steps $d\!z$. Firstly, the PP field propagation at $d\!z$ is calculated 
$$E(z,\omega)\xrightarrow{P^{(3)}\left[ E(z,\omega)\right]} \hat{E}(z+d\!z,\omega)$$ 
and then the propagated field $\hat{E}(z+d\!z,\omega)$ is used to numerically re-evaluate the $P^{(3)}_{a/b}(\omega)$ to correct the probe field propagation:
$$E(z,\omega)\xrightarrow{P^{(3)}\left[ \frac{E(z,\omega)+\hat{E}(z+d\!z,\omega)}{2}\right]} E(z+d\!z,\omega)$$

In Figure \ref{Fig:Fit_red} the experimental signal (circles) is compared with the globally fitted one (dashed lines) in the red (panels a-b) and in the blue (panels c-d) side of the spectrum by such procedure, showing a good agreement. Specifically, homogeneous lineshapes ($g(\tau)=(2\pi)^{-1}$) have been used for all the Raman modes under investigation, while the Raman pump has been modeled considering a transform limited Gaussian temporal/spectral profile (with 1.2 ps of time duration), which reflects the dependence of the SRS area (reported in Figure \ref{Fig:Area}) as a function the RP-PP temporal delay. In fact, as discussed in Appendix \ref{A: SRS Area}, such dependence can be exploited to directly determine the RP intensity temporal profile.
The cyclohexane Raman mode positions~\cite{book_Shimanouchi}  have been used to calibrate the RP wavelength, while the integration step used is 50 $\mu$m. The dephasing times $\tau_{g'_k g}$ and the relative amplitudes ($|\mu_{eg}|^2 |\mu_{eg'_{k}}|^2$) extracted from the global fit procedure are reported in Table \ref{table: SRS}.
\begin{table}[hb]
	\begin{tabular}{ |c|c|c|}
		\hline
		 Peak Position (cm$^{-1}$) & Dephasing Time & Amplitude (AU) \\
		\hline
		2853 &  1.3 ps &  0.82 \\
		\hline
		2890 &  0.3 ps & 0.64 \\
		\hline
		2922 &  1.3 ps & 0.41 \\
		\hline
		2938 &  0.7 ps  & 1.0  \\
		\hline
	\end{tabular}
	\caption{Peak positions ($\tilde{\nu}_{g'_kg}$), dephasing times ($\tau_{g'_k g}$) and relative amplitudes ($|\mu_{eg}|^2 |\mu_{eg'_{k}}|^2$) of the cyclohexane high-frequency modes, obtained from the global fit of the SRS spectra.}\label{table: SRS}
\end{table}
The unique experimental control knob for tuning the effective spectral resolution is the time delay between the Raman and the probe field~\cite{Yoon2005}: upon the stimulation of the vibrational coherences at the arrival time of the probe pulse (Figure \ref{Fig:diagrams}), the time window for probing the Raman excitations is determined by the residual temporal envelope of the Raman pulse and by the vibrational dephasing time. For this reason large (positive) time delays between the RP-PP pair can help to increase the spectral resolution, at the expense of a decreased peak amplitude.

The modeled signal is able to capture the dependence of the SRS spectra both on the RP-PP temporal delay and, most importantly, on the RP fluence. 
In particular, under the large light-matter interaction limit recorded at the highest RP energy, the SRS gain becomes exponential and the Raman profiles undergoes a spectral narrowing in the red side of the spectrum (as opposed to an exponential decrease and a broadening in the blue side). 
It is worth to stress that, while the natural width of the measured modes appears narrower under the high fluence regimes, the effective spectral resolution (intended as the capability of distinguishing different modes) of the linear gain regime is maintained, since the overall spectral profile is undergoing the same exponential gain, and not the different spectral components~\cite{PRapp_2020}. In fact, as exemplified in the monochromatic RP limit, the total RG is
$
RG(\omega)=e^{-\frac{2\pi \omega_{P} \Im \left[\sum_k \chi^{(3)}_{k} \right] I_R z}{c n_{P} }}
$,
which does not correspond to the sum of the different Raman profiles
$
\sum_k e^{-\frac{2\pi \omega_{P} \Im \left[ \chi^{(3)}_{k} \right] I_R z}{c n_{P} }}
$. Notably, under the low fluence regime, 
the expansion of the $RG$ to the first order in $I_R z$ 
leads to a signal that corresponds to the usual sum of the individual Raman profiles, equal to the spontaneous Raman response. 

\section{Conclusions}

In summary, we have studied the stimulated Raman scattering response in the large light-matter interaction limit. Measuring the SRS spectrum of liquid cyclohexane in the 
$\mathrm{C-H}$ bond vibrations spectral region (2800-3000 cm$^{-1}$), we have shown that for large RP fluences and long interaction regions of the pulses inside the sample, an exponential gain is reached in the red side of the spectrum, as opposed to an exponential decrease in the blue side. 
Under such regime, the red-side (blue-side) spectral lineshapes appear narrower (broader) with respect to the linear gain regime, while the effective resolution, relevant for spectroscopic applications, is maintained. 
Building on a perturbative treatment of the signal generation, we have shown how to model the data, taking into account for the pulse fluences, temporal profiles and relative time delays.
We anticipate that the presented results can be exploited for the rational design of novel optical devices based on SRS, such as Raman-based lasers, frequency shifter and Raman amplifier.

\appendix
\section{Dependence of the SRS Area on the Raman pump temporal profile}\label{A: SRS Area}

An electronically non-resonant SRS experiment performed under the low fluence regime can be exploited for directly accessing the RP temporal profile, by measuring the area of the Raman profiles as a function of the RP-PP temporal delay. For convenience, the expression of the red-side third-order polarization in Eq.\ref{Eq:Pta} can be recast by switching the integration variables from the time intervals $\tau_n$ to the absolute times at which the interactions with the fields occur, \textit{i.e.} $\tau_n = t_{n+1} - t_n$. By considering an impulsive probe pulse arriving at $t=T$, \textit{i.e.} $E_P(t)= \mathcal{A}^0_P \, 2 \pi \delta(t-T)$, and a Raman pump field $E_R(t)$ whose envelope is centered around $t=0$, it holds

\begin{multline}
    P^{(3)}_a(\omega)\propto (i)^3
    \sum_k
    \int_{-\infty}^{\infty} dt \, e^{i\omega t}
    \int_{-\infty}^{t}dt_3\,  E_R(t_3) \\
    \int_{-\infty}^{t_3}dt_2\, \mathcal{A}^0_P \, \delta(t_2-T)
    \int_{-\infty}^{t_2}dt_1\, 
    E^*_R(t_1) 
    \\
    e^{-i\tilde{\omega}_{ge}(t_2-t_1)}
    e^{-i\tilde{\omega}_{gg'_k}(t_3-t_2)}
    e^{-i\tilde{\omega}_{eg'_k}(t-t_3)}
\end{multline}

and hence, by adding the proper Heaviside step function in order to simplify the Dirac delta, the last equation reads 

\begin{multline}
    P^{(3)}_a(\omega)\propto (i)^3 \mathcal{A}^0_P
    \sum_k
    \int_{-\infty}^{\infty} dt \, e^{i\omega t} \\
    \int_{-\infty}^{t}dt_3\, \theta(t_3-T) \,
    E_R(t_3)\,  e^{-i\tilde{\omega}_{eg'_k}(t-t_3)} \, e^{-i\tilde{\omega}_{gg'_k}(t_3-T)}\\
    \int_{-\infty}^{T}dt_1\, 
    E^*_R(t_1) \,  e^{-i\tilde{\omega}_{ge}(T-t_1)}
\end{multline}

Under non-resonant conditions, the dominant contributions from the integrals over $t_3$ and $t_1$ come for $t_3 = t$ and $t_1 = T$, hence
$$
    P^{(3)}_a(\omega)\propto -i
    \int_{-\infty}^{\infty} dt \, e^{i\omega t} \theta(t-T) \, E_R(t) \,
    E^*_R(T) \,
    e^{-i\tilde{\omega}_{gg'_k}(t-T)}
$$
where the summation over the different normal modes has been omitted for simplicity.
Under the low fluence regime, the area $A$ of the SRS profiles can be directly obtained by integrating the heterodine-detected signal $ RG(\omega)-1 = S(\omega) \propto - \Im \left( E^*_P(\omega) P^{(3)}_a(\omega)/|E_P(\omega)|^2\right)$ over the detection frequency $\omega$. Since $E^*_P(\omega) = \mathcal{A}^0_P e^{i \omega T}$, it holds 
\begin{equation}
	\begin{gathered}
    A= - \int_{-\infty}^{\infty} d\omega \,\,
    \Im\left(E^*_P(\omega)
    P^{(3)}_a(\omega)\right) \propto 
    \\
    \Re \Big(
    \int_{-\infty}^{\infty} d\omega \, e^{i \omega T} 
    \int_{-\infty}^{\infty} dt \, e^{i\omega t} \\ E_R(t) \,
    \theta(t-T) \,
    E^*_R(T) \,
    e^{-i\tilde{\omega}_{gg'_k}(t-T)}
    \Big)\propto\\
    \Re \Big(
    \int_{-\infty}^{\infty} dt \,
    2 \pi \delta(t-T)\, E_R(t) \,
    \theta(t-T)\,
    E^*_R(T)\,
    e^{-i\tilde{\omega}_{gg'_k}(t-T)}
    \Big)\propto\\
    \Re \Big(E_R(T)   E^*_R(T)
    \Big)= I_R(T)
	\end{gathered}
\end{equation}
which is the intensity of the Raman pump at the arrival time of the probe pulse. Similar considerations can be derived also for the blue side of the SRS spectrum.
Hence, the measurement of the area under the SRS bands as a function of the RP-PP delay directly provides the temporal profile of the Raman pump intensity.

\noindent\rule{7cm}{0.4pt}

\section{Acknowledgement}
This project has received funding from the PRIN 2017 Project, Grant No.
201795SBA3-HARVEST, and from the European Union’s Horizon 2020 research and innovation program Graphene Flagship under Grant Agreement No. 881603.
G.B. and T.S. acknowledge the `Progetti di Ricerca Medi 2019' and the `Progetti di Ricerca Medi 2020' grants by Sapienza~Universit\'a~di~Roma.

\bibliographystyle{unsrt}

\bibliography{biblio6}

\end{document}